\documentclass[aps,pra,twocolumn,superscriptaddress,showkeys,showpacs]{revtex4-1}
\usepackage[T1]{fontenc}
\usepackage{standalone}
\usepackage{amsthm,amsfonts,amstext,amsmath,amssymb}
\usepackage{nccfoots}
\usepackage{comment}
\usepackage{graphicx}
\usepackage{braket}
\usepackage{xcolor}
\usepackage{placeins}
\usepackage{verbatim}
\usepackage{todonotes}
\usepackage[makeroom]{cancel}

\definecolor{mygreen}{rgb}{0,0.5,0}
\definecolor{myblue}{rgb}{0,0,0.75}
\definecolor{mymagenta}{cmyk}{0,1,0,0.12}
\usepackage[colorlinks=true,
   linkcolor=blue,
   citecolor=blue,
   urlcolor =blue]{hyperref}

\newcommand{\citeSM}{\cite[{\tiny SM}\kern-0.3em][]{SM}}

\newcommand{\be}{\begin{equation}}
\newcommand{\ee}{\end{equation}}

\newcommand{\tr}[2][]{\text{Tr}_{#1}\left[#2\right]}
\newcommand{\probs}[2][]{\ensuremath{\left\langle \mathrm{P}\left(#2\right)^{#1} \right \rangle}}
\newcommand{\probsU}[2][]{\ensuremath{\mathrm{P}\left(#2\right)^{#1} }}
\renewcommand{\vec}[1]{\mathbf{#1}}
\newcommand{\ketbra}[2]{\ensuremath{\ket{#1}\bra{#2}}}
\newcommand{\hilbertn}{\mathcal{N}_A}

\newcommand{\hilberta}{\mathcal{N}_A^{(\balpha)}}
\newcommand{\balpha}{{\boldsymbol{\alpha}}}
\newcommand{\CUEH}[2]{\text{CUE}\left(\hilbertn\right)}

\expandafter\let\csname equation*\endcsname\relax

\expandafter\let\csname endequation*\endcsname\relax

\usepackage{amsmath}

\begin{document}
\title{Unitary $n$-designs via random quenches in atomic Hubbard and Spin models: Application to the measurement of R\'enyi entropies}

\author{B. Vermersch}
\thanks{These two authors contributed equally.}
\affiliation{Institute for Theoretical Physics, University of Innsbruck, Innsbruck,
Austria}
\affiliation{Institute for Quantum Optics and Quantum Information, Austrian Academy
of Sciences, Innsbruck, Austria}

\author{A. Elben}
\thanks{These two authors contributed equally.}
\affiliation{Institute for Theoretical Physics, University of Innsbruck, Innsbruck,
Austria}
\affiliation{Institute for Quantum Optics and Quantum Information, Austrian Academy
of Sciences, Innsbruck, Austria}

\author{M. Dalmonte}
 \affiliation{International Center for Theoretical Physics, 34151 Trieste, Italy}

\author{J. I. Cirac}
\affiliation{Max-Planck-Institut f\"ur Quantenoptik, Hans-Kopfermann-Str. 1, D-85748 Garching, Germany}

\author{P. Zoller}

\affiliation{Institute for Theoretical Physics, University of Innsbruck, Innsbruck,
Austria}
\affiliation{Institute for Quantum Optics and Quantum Information, Austrian Academy
of Sciences, Innsbruck, Austria}
\affiliation{Max-Planck-Institut f\"ur Quantenoptik, Hans-Kopfermann-Str. 1, D-85748 Garching, Germany}
\begin{abstract}
We present a general framework for the generation of random unitaries based on random quenches in atomic Hubbard and spin models, forming approximate unitary $n$-designs,  and their application to the measurement of R\'enyi entropies. 
We  generalize our protocol presented in Ref.~\cite{Elben2018} to a broad class of atomic and spin lattice models. We further present an in-depth numerical and analytical study of experimental imperfections, including the effect of decoherence and statistical errors, and discuss connections of our approach with many-body quantum chaos.

\end{abstract}
\date{\today}

\maketitle

\makeatletter
\def\l@subsubsection#1#2{}
\def\l@subsection#1#2{}
\makeatother

\section{Introduction}

In a recent paper~\cite{Elben2018} we proposed a protocol to measure R\'enyi entropies of atomic Bose and Fermi Hubbard~\cite{Murmann2015,Haller2015,Greif2016,Boll2016,Cheuk2016,Labuhn2016,Zeiher2017,Bernien2017} and spin systems~\cite{Jurcevic2017,Zhang2017}. The aim was to develop a measurement scheme, which can be implemented in state of the art AMO~\cite{Bloch2012,Blatt2012,Browaeys2016} and solid-state experiments~\cite{Houck2012,Cai2013}. The key ingredient was the realization of random measurements \cite{VanEnk2012} on a quantum many-body system of interest. This random measurement was realized as a two step procedure: First, a series of random quenches was applied to the subsystem of the quantum many-body system of interest, $A\subset\mathcal{S}$. Second, performing single shot readouts with a quantum gas microscope~\cite{Kuhr2016}, allowed one to obtain the R\'enyi entropy of the subsystem $A$. In contrast to previous proposals ~\cite{Daley2012,Abanin2012} and experiments~\cite{Islam2015,Kaufman2016}, where R\'enyi entropies could be extracted from identical {\em copies} of quantum systems, the present protocol is realized with a {\em single} quantum system and is therefore immediately implementable in present 1D and 2D atomic experiments with quantum gas microscopes.

While Ref.~\cite{Elben2018} outlined the basic protocol illustrated by applications, including the entropy growth in the many-body localized phase ~\cite{Basko2006,Bardarson2012,Serbyn2013,Nandkishore2015,Altman2015} and the measurement of an area law of entanglement  in a 2D system~\cite{Song2011}, it is the aim of the present paper to elaborate on  details behind the measurement scheme and extend the list of applications to model cases of interest. Specifically, this includes a detailed numerical study of the generation of random unitaries allowing the measurement of higher ($n$-th) order R\'enyi entropies, i.e.\ realizing approximate unitary $n$-designs~\cite{Gross2007,Roy2009},  for a broad range of  atomic Hubbard  and spin models. We also analytically estimate, for arbitrary $n$, the statistical errors involved in  such a measurement and study the role of the experimental imperfections and decoherence.
Finally, we relate the time evolution under random quenches to the thermalization dynamics of closed systems~\cite{DAlessio2013,DAlessio2014,Ponte2015,Gopalakrishnan2016,Bordia2017,Machado2017}  and highlight close connections to quantum chaos~\cite{Haake2010}.

The paper is organized as follows. In Sec.~\ref{sec:protocol} we review and generalize the  protocol for measuring R\'enyi entropies based on random quenches. 
In Sec.~\ref{sec:quenches}, we provide a detailed numerical study of creation of approximate $n$-designs in atomic Hubbard and spin models.
In Sec.~\ref{sec:errors}, we discuss the influence of statistical errors and other sources of imperfections and estimate the necessary amount of experimental resources (number of measurements). Finally, in Sec.~\ref{sec:chaos}, we discuss relations of our approach to random matrix theory and many-body quantum chaos.

\section{Measurement protocol for R\'enyi entropies}\label{sec:protocol}
\begin{figure}[h]
	\includegraphics[width=0.95\columnwidth]{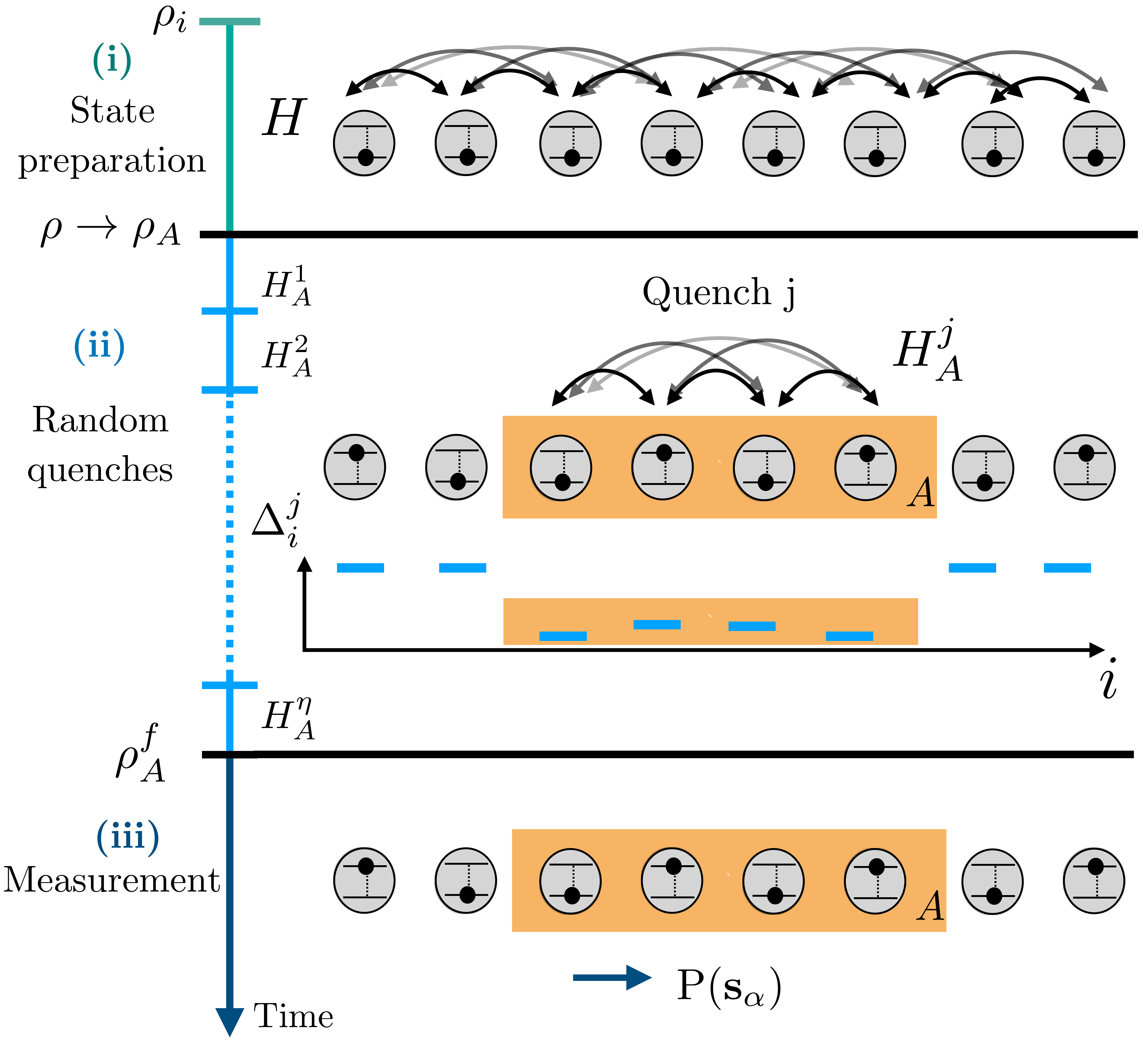}
	\caption{{\it Measuring R\'enyi entropy via random quenches.} Experimental sequence (illustrated in the context of spin chains) consisting of  (i) quantum state preparation of a many-body quantum state $\rho$, (ii) Isolation of a partition $A$ with reduced density matrix $\rho_A$ via energy offsets and random unitary evolution in $A$ via random quenches $j=1,\dots,\eta$ realized by disorder patterns $\Delta_i^j$, and (iii) a projective measurement on $A$. \label{fig:setup}}
\end{figure}

In this  section, we review, building on Refs.~\cite{Elben2018,VanEnk2012},  the protocol for the measurement of R\'enyi entropies in atomic Hubbard or spin models. Beyond Ref.~\cite{Elben2018},  we  generalize our framework to incorporate random measurements of arbitrary observables and give explicit expressions to extract R\'enyi entropies of arbitrary order $n$.

\subsection{R\'enyi entropies}

R\'enyi entropies are essential tools for the characterization of entanglement properties in correlated phases of quantum matter~\cite{Eisert2010}.
Given a  state $\rho$ in a quantum system $\mathcal{S}$, the R\'enyi entropies associated with the reduced density matrix $\rho_A=\tr[\mathcal{S}\backslash A]{\rho}$ of a subsystem $A\subset\mathcal{S}$ are for $n\geq 0$ defined as 
\begin{align}
S^{(n)}(\rho_{A})=\frac{1}{1-n}\log{\mathrm{Tr}(\rho_{A}^n)} \; .
\end{align}
In quantum simulators, an outstanding challenge is to measure R\'enyi for partitions sizes $A$ and systems, where full quantum state tomography is not available or feasible~\cite{Gross2010,Cramer2010,Ohliger2013,Lanyon2016}.
In the following,  we are interested  in R\'enyi entropies for integer values of $n \in \mathbb{N}_{>0}$, determined by  functionals $p_n\equiv \mathrm{Tr}(\rho_{A}^n)$  of the reduced density matrix $\rho_A$.

\subsection{Experimental sequence}
\label{sec:expseq}

The experimental protocol we have in mind is shown in Fig.~\ref{fig:setup} and consists of three steps: 

\textit{Step (i)} -- The quantum state $\rho$ of interest, defined in the full many-body system $\mathcal{S}$,  is 
 prepared, obtained for instance as pure state $\rho=\ket{\Phi}\bra{\Phi}$ from the quench dynamics of a many-body Hamiltonian $H$.

\textit{Step (ii)}  -- The spatial partition $A$ is isolated from the full system $\mathcal{S}$, for instance by potential offsets (c.f Fig.~\ref{fig:setup}), and a random unitary operator $U_A$  acting on the reduced density matrix $\rho_A$ of the subsystem $A$ 
$\rho_A\rightarrow U_A\rho_AU_A^\dagger$ is realized as time evolution operator
\begin{equation}
U_A= e^{-iH_A^\eta T} \cdots  e^{-iH_A^1T} \label{eq:UA},
\end{equation}
under a series $j=1,\dots, \eta$ of random quenches applied to $A$. Here, the Hamiltonian $H_A^j $ governs the dynamics during the quench $j$ of time $T$, and $T_{\rm tot}\equiv \eta T$ denotes the total time. For each quench $j$, the  Hamitonian  $H_A^j $ can be decomposed as
 \begin{equation}
 H_A^j = H|_A+\sum_{\vec{i}\in A,\sigma} \Delta^j_\vec{i,\sigma} X_{\vec{i},\sigma}. \label{eq:HA}
 \end{equation}
The first term $H|_A$ corresponds to the static Hamiltonian of an interacting Hubbard or spin model (see below), restricted to the partition $A$. The second term represents spatial disorder, which is drawn  for each quench $j$ from a probability distribution. Here, $X_{\vec{i},\sigma}$ is a local operator acting at site $\vec{i}$ with internal degree of freedom (for instance spin) $\sigma$, and $\Delta_{\vec{i},\sigma}^j$ is the disorder potential.  We discuss examples, realizable in atomic quantum simulators with current technology, in the next section.  We note that,
in the presence of  symmetries of the quench Hamiltonian evolution, $U_A$  decomposes into blocks, $U_A= \bigoplus_{\balpha} U_A^{\balpha}$, labeled by quantum numbers $\balpha=(\alpha_1,\dots,\alpha_q)$ which correspond to  a set of conserved quantities  $Q_p$ ($p=1,\dots q$),  with   $\left[ U_A, Q_{p}\right] = 0 $.  The (reduced) states $\rho_A = \bigoplus_{\balpha} \rho_A^{\balpha}$, $\rho^f_A = \bigoplus_{\balpha} \rho_A^{f\balpha}$ have the same block structure, due to the same conservation laws. 

\textit{Step (iii)} -- An observable $\mathcal{O}$ is measured by a projective measurement on the state $U_A\rho_AU_A^\dagger$. 
   Here, we  assume that the  measurement does not mix sectors corresponding to conserved quantities [see step (ii)], i.e.~must resolve the quantum numbers $\balpha$ ($\left[Q_p,\mathcal{O}\right]=0$ for all $p=1,\dots,q$). Otherwise, the observable $\mathcal{O}$ can be arbitrary.
 In the following, we denote the  outcome of such a measurement with $\mathbf{s}_{\balpha}$, where  $\balpha$ are the associated the quantum numbers,  and the corresponding  projector with $\mathcal{P}^\mathcal{O}_{\vec{s}_{\balpha}}$.
 
Finally, steps (i)-(iii)  have  to be repeated, first  using the same random unitary $U_A$, i.e.\ the same series of random quenches, to obtain estimates of the  probabilities  $ \mathrm{P}(\vec{s}_\balpha)  = \tr[A]{U_A \rho_A U_A^\dagger \mathcal{P}^\mathcal{O}_{\vec{s}_\balpha}}$ of   measurement outcomes $\vec{s}_\balpha$; and secondly for different random unitaries to obtain R\'enyi entropies from an ensemble average (see next section). We denote in the following the number of measurements per random  unitary with $N_M$ and the number of random unitaries with $N_U$.
 
 \subsection{Extraction of R\'enyi Entropies}
 \label{sec:R\'enyis}

\subsubsection{$n$-design identities }

Our goal is to infer from an  ensemble average of the outcome probabilities $ \mathrm{P}(\vec{s}_\balpha)$ of random measurements over the generated random unitaries,  the $n$-th order R\'enyi entropy $S^{(n)}(\rho_{A})$. For this, the generated unitaries $U_A^{\balpha}$ are required to realize an (approximate) unitary $n$-design; i.e.\ to form an ensemble that reproduces the ensemble average over the full circular unitary ensemble (CUE)  for  polynomials up to degree $n$ \cite{Gross2007,Roy2009}. In particular, we use, for an estimation of the purity, the  identities
\begin{align}
&\langle u_{m,n}u_{m',n'}^* \rangle = \frac{\delta_{m,m'}\delta_{n,n'}}{\hilberta}  \label{eq:1design} 
\end{align}
and 
\begin{align}
&\langle u_{m_1,n_1} u_{m'_1,n'_1}^*u_{m_2,n_2}u_{m_2',n_2'}^*  \rangle =\nonumber \\& \frac{\delta_{m_1,m_1'}\delta_{m_2,m_2'}\delta_{n_1,n_1'}\delta_{n_2,n_2'}+\delta_{m_1,m_2'}\delta_{m_2,m_1'}\delta_{n_1,n_2'}\delta_{n_2,n_1'}}{\mathcal{N}_A^{(\balpha)2}-1}\nonumber\\
&\!\! -\frac{\delta_{m_1,m_1'}\delta_{m_2,m_2'}\delta_{n_1,n_2'}\delta_{n_2,n_1'}+\delta_{m_1,m_2'}\delta_{m_2,m_1'}\delta_{n_1,n_1'}\delta_{n_2,n_2'}}{\hilberta(\mathcal{N}_A^{(\balpha)2}-1)},  \label{eq:2design} 
\end{align}
fulfilled in unitary $1$- and $2$-designs, respectively~\cite{Collins2006,Puchaa2017}. Here, $\langle\dots \rangle$ denotes the average over the ensemble of (generated) random unitaries,  $u$ the matrix elements of $U_A^\balpha$ and $\hilberta$ the dimension of the sector $\balpha$.

On a quantum computer, approximate $n$-designs can be created via random circuits which consist of random sequences of two qubit gates~\cite{Dankert2009,Ohliger2013,Brandao2016}.
In Sec.~\ref{sec:quenches}, we show that, in atomic Hubbard and spin models, random unitaries forming approximated $n$-designs can be generated via random quenches (Eq.~\eqref{eq:UA}), with an error $\epsilon$ which decreases exponentially with the number of quenches $\eta$.

\subsubsection{Purity from measurements of arbitrary observables}

Assuming the $1$- and $2$-design identities,  Eqs.~\eqref{eq:1design} and \eqref{eq:2design}, we obtain from an  average  of the outcome probabilities $ \mathrm{P}(\vec{s}_\balpha)  = \tr[A]{U_A \rho_A U_A^\dagger \mathcal{P}^\mathcal{O}_{\vec{s}_\balpha}}$ over the random unitary ensemble
\begin{align}
\probs{\vec{s}_\balpha}  = &\frac{\tr{\mathcal{P}^\mathcal{O}_{\vec{s}_\balpha}}\tr{\rho_A^{(\balpha)}}}{\hilberta} \label{eq:p1}\\
\probs[2]{\vec{s}_\balpha} = &\frac{\tr{\mathcal{P}^\mathcal{O}_{\vec{s}_\balpha}}^2\tr{\rho_A^{(\balpha)}}^2+ \tr{\mathcal{P}^\mathcal{O}_{\vec{s}_\balpha}}\tr{\rho_A^{(\balpha)2}}}{{\hilberta}^2-1} \nonumber \\ &- \frac{\tr{\mathcal{P}^\mathcal{O}_{\vec{s}_\balpha}}^2\tr{\rho_A^{(\balpha)2}}+ \tr{\mathcal{P}^\mathcal{O}_{\vec{s}_\balpha}}\tr{\rho_A^{(\balpha)}}^2}{\hilberta({\hilberta}^2-1)} \; . \label{eq:p2} 
\end{align}
These equations can be solved to determine $\tr{\rho_A^{(\balpha)2}}$,  and by summation over all sectors $p_2= \sum_{\balpha}  \tr{\rho_A^{(\balpha)2}}$, which gives access finally to $S^{(2)}(\rho_A)$.

For example, in the Fermi Hubbard (FH) model (see below)  the conserved quantities  are the total magnetization $S_z$ and total particle number $N$, i.e $\balpha=S_z,N$.
An observable resolving $S_z$ and $N$ is the spin-resolved measurement of the local particle number with a quantum gas microscope. Given an outcome $\mathbf{s}_{N,S_z}=(\vec{n}_\uparrow, \vec{n}_\downarrow)$ of such a measurement, we find $S_z= \sum_\vec{i} \vec{n}_{\uparrow,\vec{i}} - \vec{n}_{\downarrow,\vec{i}}$ and $N=\sum_\vec{i} \vec{n}_{\uparrow,\vec{i}} + \vec{n}_{\downarrow,\vec{i}}$. The projectors are $\mathcal{P}^\mathcal{O}_{\mathbf{s}_{N,S_z}}=\ketbra{\vec{n}_\uparrow, \vec{n}_\downarrow}{\vec{n}_\uparrow, \vec{n}_\downarrow}$ with the Fock states $\ket{\vec{n}_\uparrow, \vec{n}_\downarrow}$ and $\tr[]{\mathcal{P}^\mathcal{O}_{\mathbf{s}_{N,S_z}}}=1$.

\subsubsection{Higher order functionals $\tr[]{\rho_A^n}$}

We describe now the estimation of higher order functionals $p_n$, related to averages of higher order polynomials of outcome probabilities $\probs[n]{\vec{s}_\balpha} $. We adopt definitions and notations from the previous subsection. For measurements of  observables with $\tr{\mathcal{P}^\mathcal{O}_{\vec{s}_\balpha}}>1$,  the
ensemble averages $\probs[n]{\vec{s}_\balpha} $ ($n\in \mathbb{N}$) can be evaluated order by order in $n$ using the Weingarten calculus \cite{Collins2006,Puchaa2017} for $n$-designs. An explicit formula can be given for direct measurements of occupation probabilities of basis states ($\tr{\mathcal{P}^\mathcal{O}_{\vec{s}_\balpha}}=1$) based on Isserlis' theorem \cite{VanEnk2012,ISSERLIS1918}.  The evaluation of $ \probs[n]{\vec{s}_\balpha} $ reduces then  to the counting of permutations in the symmetric group $S_n$. One finds
\begin{align}
\langle  \probsU[n]{\vec{s}_\balpha} \rangle \stackrel{\tr{\mathcal{P}^\mathcal{O}_{\vec{s}_\balpha}}=1}{=}  \frac{1}{D_n} \sum_{\substack{  {b_1,\dots,b_n \in \mathbb{N}_0}\\ {\text{with}} \\ {\sum_{l=1}^n l b_l = n}  } } C_{b_1,\dots, b_n} \prod_{k=1}^n \text{Tr}\left[\rho_A^{(\balpha)k} \right]^{b_k} 
\label{eq:higher}
\end{align}
where $D_n= \prod_{i=0}^{n-1} ( \hilberta +i) $. $C_{b_1,\dots, b_n}$  denotes the number of  permutations $\pi \in S_n$ with $\operatorname {typ} (\pi )=1^{b_{1}}2^{b_{2}}\ldots n^{b_{n}}$  and is given by \cite{Bona2016} 
\begin{align}
C_{b_1,\dots, b_n} = {\frac  {n!}{b_{1}!\cdot b_{2}!\cdot \ldots \cdot b_{n}!\cdot 1^{{b_{1}}}\cdot 2^{{b_{2}}}\cdot \ldots \cdot n^{{b_{n}}}}} \; .
\end{align}
Knowing $\text{Tr}\left[\rho_A^{(\balpha)k} \right]^{b_k} $ for $k<n$, Eq.~\eqref{eq:higher} can be solved to obtain $\tr{\rho_A^{(\balpha)n}}$. By summation over all sectors, one finds  $\tr[]{\rho_A^n}= \sum_{\balpha}  \tr{\rho_A^{(\balpha)n}}$.

\section{Creation of $n$-designs via random quenches}\label{sec:quenches}

\subsection{General strategy}

In the following, we present  a numerical study of the generation of approximate unitary $n$-designs via random quenches in various atomic Hubbard and spin models. We extend the results on $2$-designs presented in Ref.~\cite{Elben2018} for the one- and two-dimensional Heisenberg, by an analogous study in Quantum Ising and Fermi- and Bose Hubbard models and in particular give  the details on the creation of higher order designs and optimization of parameters.
To test the $n$-design properties of the generated random unitaries, we choose test states $\rho_A$ and compare the estimated functional $(p_n)_e\equiv \left(\tr{\rho_A^n}\right)_e$ (i.e.~obtained using the protocol described in Sec.~\ref{sec:protocol} with the generated set of random unitaries) with the exact value $p_n=\tr{\rho_A^n}$. In all models, we consider  partitions $A$ consisting of $L$ sites. For two-dimensional models, these partitions are rectangular, with $L=L_x\times L_y$.

\subsection{The 1D Quantum Ising model and the generation of higher-order designs}\label{app:Ising}

Here,  we study  the generation of random unitaries in the Quantum Ising model, which has been implemented with Rydberg atoms~\cite{Labuhn2016,Zeiher2017,Bernien2017} and trapped ions~\cite{Jurcevic2017,Zhang2017}.
We consider a uni-dimensional chain of $L$ spins, where the quenches are governed by Hamiltonians
\begin{eqnarray}
H|_A &=& \sum_{(i<l) \in A} \frac{C_\alpha}{a^\alpha|i-l|^\alpha} \sigma_i^u \sigma_l^u + \Omega \sum_{i\in A} \sigma_i^x
\end{eqnarray}
with $a$ the lattice spacing and $u=x$, $0<\alpha<3$ ($u=z$, $\alpha=6$) for the trapped ions (Rydberg atoms) case, respectively. In the following, $J=C_\alpha/a^\alpha$ refers to the nearest neighbor coupling. The disorder term is obtained with $X_i=\sigma_i^z$, and the 
disorder potentials are drawn from a normal distribution with standard deviation $\delta$.

\begin{figure}[h!]
	\includegraphics[width=0.99\columnwidth]{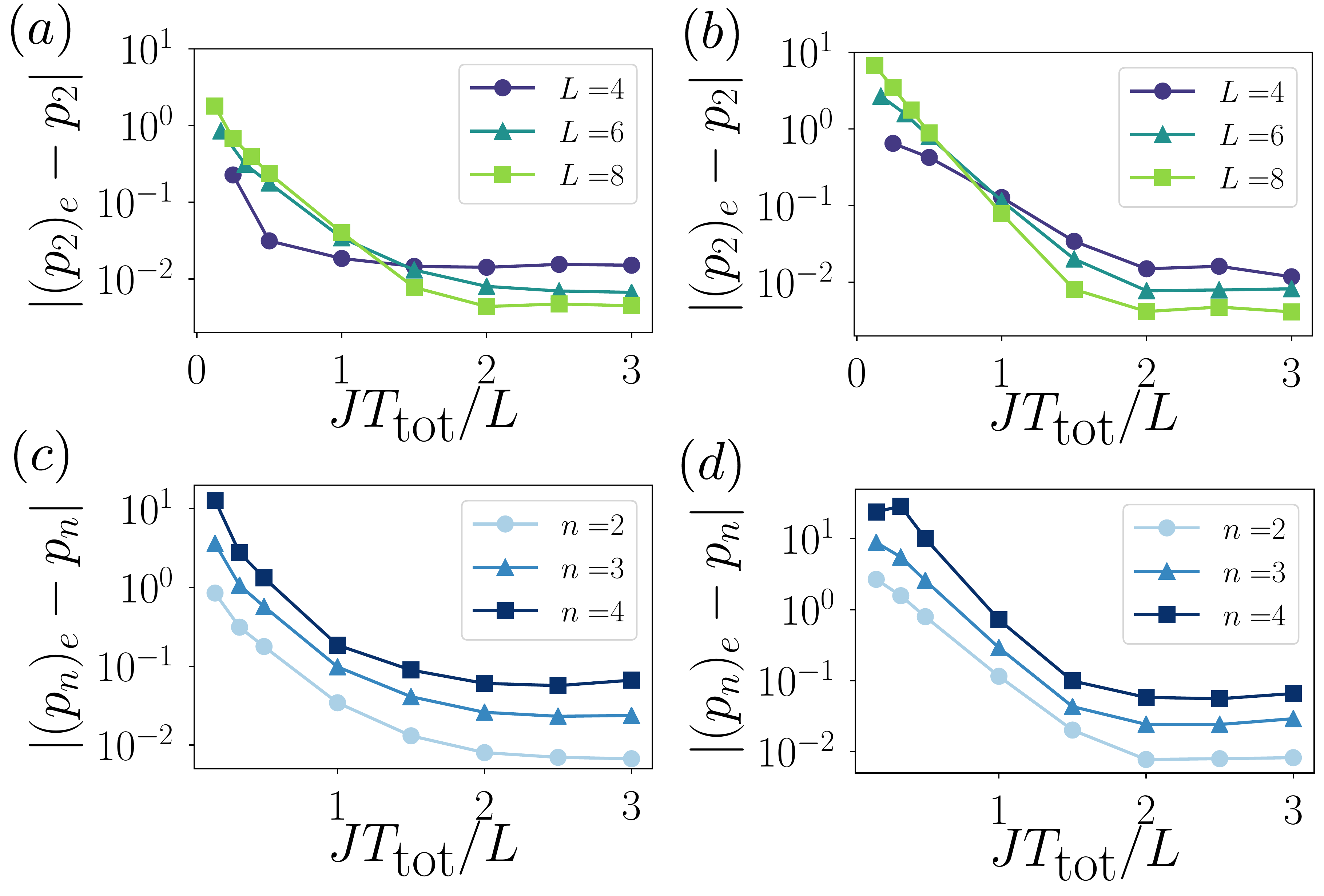}
	\caption{\textit{Creation of $n$-designs in the 1D Quantum Ising model}. 
	(a-b) Convergence to $2$-designs for various number of spins $L$ in (a) the ions case $\alpha=1.5$, $u=x$ and  (b)  the Rydberg case $\alpha=6$, $u=z$ as a function  $JT_{\textrm{tot}}/L=\eta/L$. (c-d)
	Convergence to $n=2,3,4$ designs in a system with  $L=6$ spins in (c) the ions case $\alpha=1.5$, $u=x$ and  (d)  the Rydberg case $\alpha=6$, $u=z$ as a function  $JT_{\textrm{tot}}/L=\eta/L$.
		In all panels, we use $\Omega=\delta=J=1/T$, $N_U=500$, and  $N_M=\infty$.
	 \label{fig:Ising1}}
\end{figure}

In Fig.~\ref{fig:Ising1}, panels (a-b), we show, for a fixed quench time $JT=1$, the average error of the estimated purity $(p_2)_e$  as a function of $JT_{\textrm{tot}}/L=\eta/L$, in a system consisting of $L=4,6,8$ spins.  In  panels (c-d), we investigate the average error of the estimated higher order functionals $(p_n)_e$  ($n=2,3,4$) as a function of $JT_{\textrm{tot}}/L=\eta/L$, for a fixed quench time $JT=1$ and system size $L=6$. In all cases we use here a pure antiferromagnetic test state and  observe an exponential decrease of the error of an estimate $(p_n)_e$ ($n=2,3,4$) with the number of quenches with $JT_{\textrm{tot}}/L=\eta/L$, towards the statistical error threshold (which depends on $n$ and $L$, see below). We find similar behavior and scalings for other test states, indicating the convergence to approximate $n$-designs. 
While the time to create approximate $n$-designs with fixed error rises with $n$, we note that for these low order designs $n=2,3,4,..$, it is compatible with typical AMO experimental timescales.

\begin{figure}[h!]
	\includegraphics[width=0.99\columnwidth]{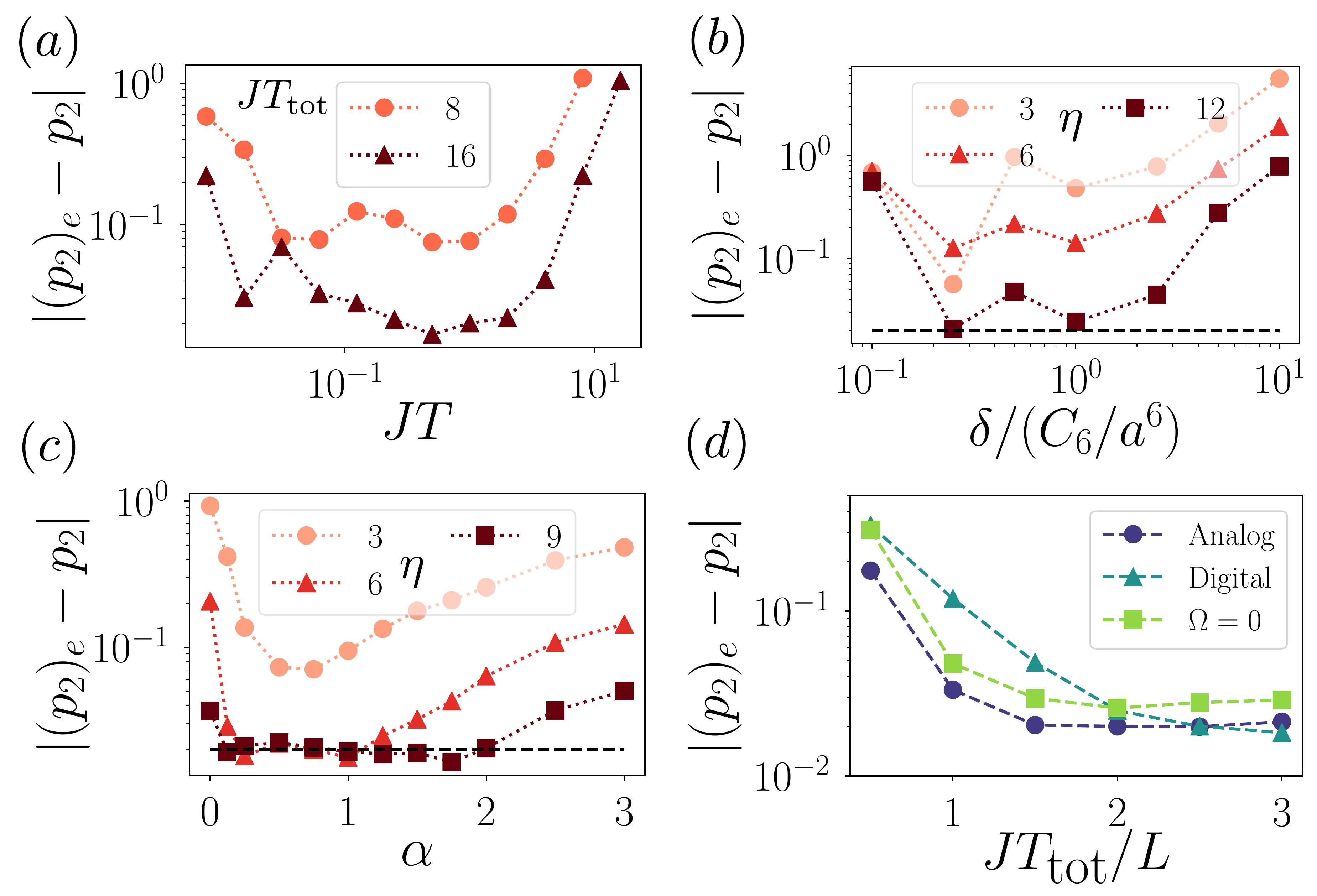}
	\caption{\textit{Optimizations  in the 1D Quantum Ising model}. 
	Optimizations of (a) the quench time $T$, (b) the disorder strength $\delta$ and (c) the power law exponent $\alpha$.
	(d) Comparison of different approaches (see text) for the quench evolutions. 
	Unless stated otherwise, we consider $\alpha=3$, $u=x$, $\Omega=\delta=J$, $L=6$, $JT= 1$, a number of randum unitaries $N_U=100$, and a number of projective measurements per unitary $N_M=\infty$.
  \label{fig:Ising2}}
\end{figure}

We discuss now the optimization of the quench parameters, to achieve maximum speed of convergence towards unitary $n$-designs. In Fig.~\ref{fig:Ising2}, 
 panels (a-b), we study, for a fixed total time $T_{\textrm{tot}}$,
the optimization of the quench parameters $T$, $\delta$. We find similar behavior as in the Heisenberg case~\cite{Elben2018}, with optimal values $J\approx \delta \approx 1/T$ when the different energy scales compete. We note that there is also an optimal value for the power law exponent $\alpha$ of the interaction, c.f.~panel (c). This can be understood as a trade-off between low-connectivity at large $\alpha$ and the emergent symmetry at $\alpha\to0$, where the interaction part of the Hamiltonian can be rewritten in terms of the collective spin operators~\cite{Lipkin1965}.
In panel (d), we show that $2$-designs can also be created (i)  in a `digital' approach  where the disorder patterns and the interactions (the transverse field  $\Omega$) are applied only on even (resp.~odd) quenches $j$, and (ii) without transverse field $\Omega=0$. Note that in the case (ii) the parity $\textrm{mod}(S_z,2)$ is a conserved quantum number.

\subsection{The 2D Fermi-Hubbard model}\label{app:FH}

In the 2D Fermi-Hubbard (FH) model, we consider the static Hamiltonian
\begin{equation}
H|_A= - t_F\sum_{\langle \vec{i,l}\rangle \in A,\sigma}c_{\vec{i}\sigma}^{\dagger}c_{\vec{l}\sigma}+U \sum_{\vec{i}\in A}n_{\vec{i}\uparrow}n_{\vec{i}\downarrow}.\label{eq:FH}, 
\end{equation}
with $t_F$ the hopping and $U$ the interaction strength. Here
 $c_{\vec{i},\sigma}^{(\dagger)}$ denote fermionic annihilation (creation)
operators at lattice site $\vec{i}=(i_x,i_y)$ and spin $\sigma\in\left\{ \uparrow,\downarrow\right\} $, and $n_{\vec{i}\sigma} = c_{\vec{i}\sigma}^{\dagger}c_{\vec{i}\sigma}$. The operators associated with the disorder patterns $\Delta_\vec{i}^j$ are (spin-dependent) chemical potentials $X_{\vec{i},\sigma}=n_{\vec{i},\sigma}$.

\begin{figure}[t]
	\includegraphics[width=0.99\columnwidth]{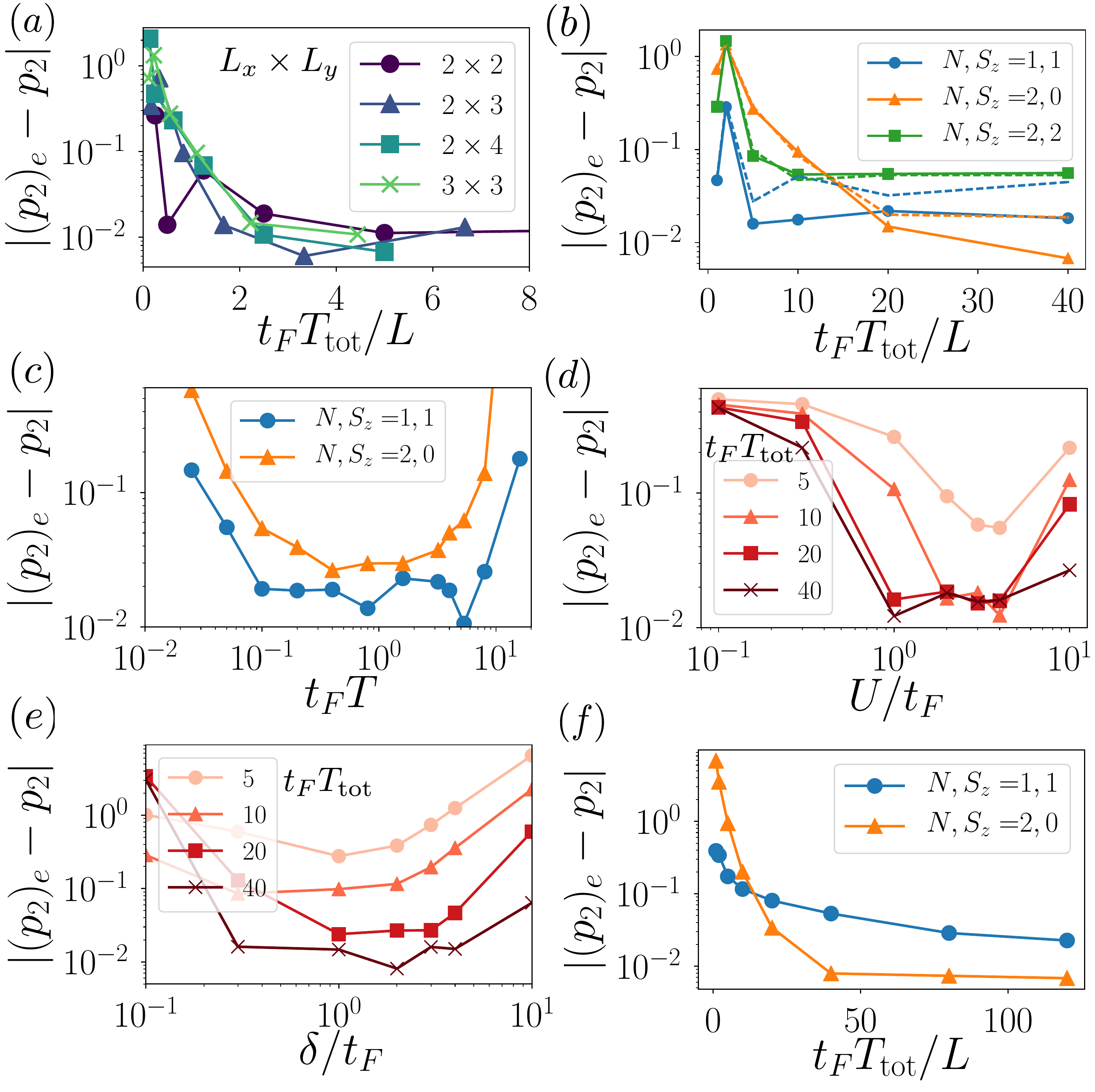} \caption{ \textit{Creation of random unitaries in the 2D Fermi Hubbard model}.
	(a) The average error of the estimated purity of $\ket{\psi_{2,0}}$ is shown as a function of $\eta$ and for different partition
			sizes $L_{x},L_{y}$. The convergence to a unitary $2$-design is exponential with scaling parameter $t_F T_\mathrm{tot}/L$.
	 (b) The error
			is shown for $N_{U}=100$ (dashed) and $N_{U}=500$ (solid lines)
			and for different sectors.
			 For $N,S_z=2,2$, this plateau is due to the
			non-interacting character of the sector (see text). 
	                (c) For  $T_\mathrm{tot}=16/t_F$, error as a function of $T$ , with an optimum at $t_F T\approx1$.	
			(d)-(e) Influence of the Hamiltonian parameters $U$ and $\delta$ on the error of the estimated purity for $\ket{\psi_{2,0}}$.
			(f) Convergence to a unitary $2$-design with  a single disorder pattern and random gate times $T_{j}$.
			Unless stated otherwise, $N_U=500$, $N_M=\infty$, $\delta=U=t_F$, $t_F T=1$ and $(L_x,L_y)=(3,3)$.\label{fig:FH} }
\end{figure}

For simplicity, we consider density matrices $\rho_A$ as pure states $\rho_A=\ket{\psi_{N,S_z}}\bra{\psi_{N,S_z}}$ of a single quantum number sector $(N,S_z)$~\footnote{We observed that the convergence to $2$-designs obeys the same scalings for mixed states}. For the examples presented in Fig.~\ref{fig:FH}, we use

\begin{eqnarray}
\ket{\psi_{1,1}} &=& (c^\dagger_{\vec{i}=(1,1),\uparrow}+c^\dagger_{(L_x,L_y),\uparrow})/\sqrt{2}\ket{V} \nonumber \\
\ket{\psi_{2,0}} &=& (c^\dagger_{(1,1),\downarrow}c^\dagger_{(L_x.L_y),\uparrow}+c^\dagger_{(L_x,1),\downarrow}c^\dagger_{(1,L_y),\uparrow})/\sqrt{2}\ket{V} \nonumber \\
\ket{\psi_{2,2}} &=& (c^\dagger_{(1,1),\uparrow}c^\dagger_{(2,1),\uparrow})+c^\dagger_{(L_x-1,L_y),\uparrow}c^\dagger_{(L_x,L_y),\uparrow})/\sqrt{2}\ket{V} \nonumber ,\\
\end{eqnarray}
with $\ket{V}$ the vacuum state. We use spin-dependent disorder potentials with $\Delta_{\bf{i},\uparrow}=2\Delta_{\bf{i},\downarrow}$, which however  differ only by a global prefactor.
Here, $\Delta_{\bf{i},\downarrow}$ is drawn from a normal distribution of standard
deviation $\delta$.

The error of the estimated purity, shown in Fig.~\ref{fig:FH}(a) for $(N,S_z)=(2,0)$ and for different 2D partitions, decreases exponentially with
the preparation time $T_\mathrm{tot}$, reaching the statistical error threshold
at $t_F T_\mathrm{tot}\sim2L$. 
In panel (b), we represent the error for different number of unitaries $N_U$ and different quantum number sectors $(N,S_z)$.
Note that the plateau associated with the sector $N=S_z=2$ does not decrease with $N_U$ and is thus not due to statistical errors. 
It originates from the fact that the FH model is a non-interacting model for this specific sector where all fermions have the same spin and thus `entangling' random unitaries
cannot be created. However, for the physical examples we are interested
in (for example Fig.~1(b) in Ref.~\cite{Elben2018}), this sector is only marginally populated leading to negligible errors in the estimated purity.
In panel (c), the error of the purity is shown as a function of the quench time $T$, while keeping
the total time $T_{\text{tot}}=16/t_F$ fixed. As in the Heisenberg model~\cite{Elben2018}, the optimal quench time $T t_F^{-1} \sim 1$ is of the order of the hopping time.
In panel (d-e), we show the optimization of the other quench parameters, the interaction $U$ and disorder $\delta$ strengths, and find optimal values being also of the order of the hopping $t_F$.

We show in Fig.~\ref{fig:FH}(f) the error of the  purity estimated with random unitaries, which are created using  a single
disorder pattern which is switched on and off after random times $T\to T_{j}$,
drawn from a uniform distribution in $[0,2/t_F]$. Note that convergence to a unitary 2-design requires a larger number of
quenches compared to the case of multiple disorder patterns [c.f panel (a)].

\subsection{The Bose-Hubbard model}

\begin{figure}
	\includegraphics[width=0.99\columnwidth]{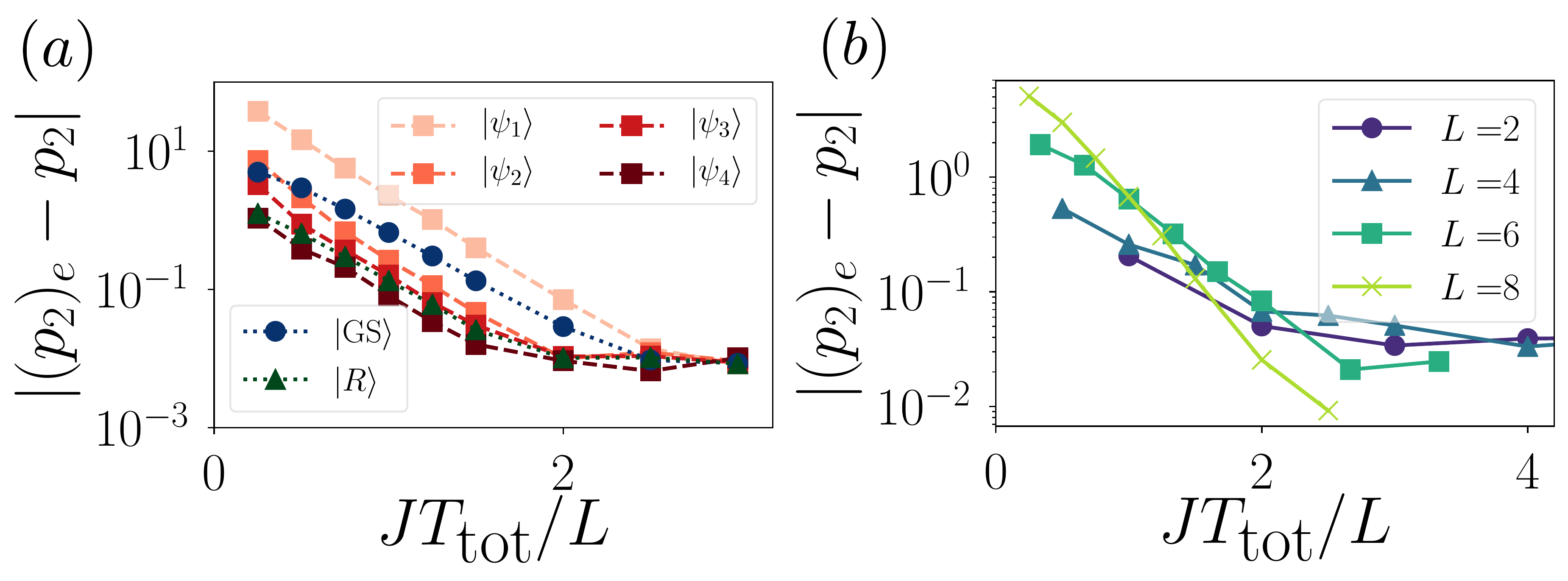}
\caption{{\it Convergence to $2$-designs in the BH model.} (a) Average error of the estimated purity $|(p_2)_e-p_2|$ for a uni-dimensional partition with $L=8$ sites and $N=4$ particles and various test states: Fock states $\ket{\psi_1}\propto (a_1^\dagger)^4\ket{V}$,$\ket{\psi_2}\propto (a_4^\dagger)^4\ket{V}$,$\ket{\psi_3}\propto a_1^\dagger a_2^\dagger a_3^\dagger a_4^\dagger\ket{V}$,$\ket{\psi_4}\propto a_1^\dagger a_3^\dagger a_5^\dagger a_7^\dagger\ket{V}$ (squares, light to dark red/gray), ground state $\ket{\psi_\mathrm{GS}}$ of the BH model with $U=J$ (circles) and vanishing disorder, random state $\ket{\psi_\mathrm{R}}$ (triangles). (b) Average error for various uni-dimensional partition sizes $L=2,4,6,8$ at half filling. Test state is the ground state of the corresponding BH Hamiltonian with $U=J$ and vanishing disorder. In both panels,  we consider $N_U=100$, $N_M=\infty$. \label{fig:BH}}
\end{figure}

Finally, in the unidimensional Bose-Hubbard (BH) model with hopping $J$ and onsite interaction $U$
\begin{equation}
H|_A=-J\sum_{i\in A}\left(a_{i+1}^{\dagger}a_{i}+\text{h.c.}\right)+\frac{U}{2}\sum_{i\in A}n_{i}(n_{i}-1),
\end{equation}
random quenches are created via onsite potentials $\Delta_i$ ($X_i=n_i$) drawn  for each quench independently from a normal distribution with standard deviation $\delta$.  Here,  $a_i$ ($a_i^\dagger$) denote bosonic annihilation (creation) operators and $n_i=a_i^\dagger a_i$ the local particle number operators.  In Fig.~\ref{fig:BH}, we show the error of the estimated purity for various uni-dimensional partitions with $L$ sites and various test states with $N=L/2$ particles (defined in the caption).  We observe, for fixed quench time $JT=1$ and $\delta=U=J$, an exponential decrease of the error with growing $JT_{\textrm{tot}}/L=\eta/L$ towards a plateau corresponding to statistical errors (see below). As in the other models, the time $JT_{\textrm{tot}}$ to create approximate $2$-design scales thus, at half filling, approximately linearly with the  number of sites $L$, with additional corrections due to the growing number of particles $N=L/2$. 

\section{Statistical errors and imperfections}\label{sec:errors}

The goal of this section is to present a detailed study of statistical errors and the influence of possible experimental imperfections, such as decoherence, finite detection fidelity and limited disorder reproducibility, on the measurement of R\'enyi entropies via the presented protocol.

\subsection{Statistical errors}

We first discuss statistical errors involved in the protocol presented in section \ref{sec:protocol}. Hereby, we support the numerical findings presented in Ref.~\cite{Elben2018} with analytical considerations, and extend to higher order R\'enyi entropies. We aim to determine the necessary experimental resources,  the number of  measurements $N_M$ per random unitary and  number of random unitaries $N_U$, to estimate functionals $p_n=\tr[]{\rho_A^n}$ ($n\in \mathbb{N}$)  of a density matrix $\rho_A$ up to a given error threshold and the scaling of these errors with system size.

To determine the statistical errors, we assume  that the reduced density matrix $\rho_A$ of the subsystem $A$ is defined on a Hilbert space  with dimension $\hilbertn$ in which random unitaries $U_A \in \CUEH{A}{}$ (or more generally from an exact $n$-design) can be created.   We consider further first random measurements of an observable $\mathcal{O}$ with $N_\mathcal{O}=\hilbertn$ possible outcomes (i.e.\  $\tr{\mathcal{P}^\mathcal{O}_{\vec{s}}}=1 \; \forall \vec{s}$ where $\mathcal{P}^\mathcal{O}_{\vec{s}}$ denote the  projector corresponding to a measurement outcome $\mathbf{s}$).  We then extend to more general observables and discuss additionally the influence of the  state $\rho_A$ itself on the statistical errors.

\begin{figure}[t]
	\includegraphics[width=0.89\columnwidth]{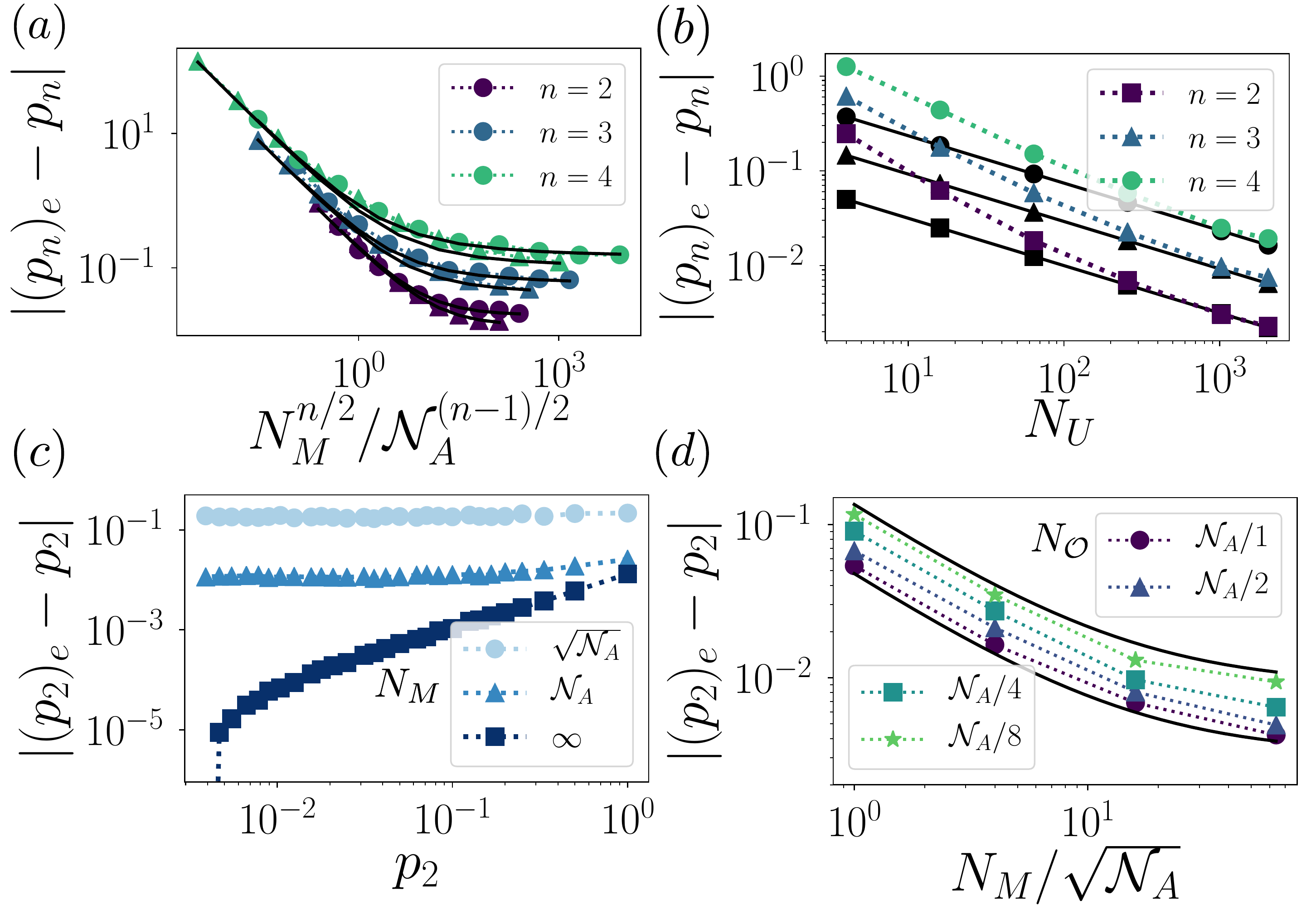}	
	\caption{\textit{Scaling of statistical errors}. Average
		statistical error of an estimation of $p_n=\tr[]{\rho_A^n}$ as a function of a) the number of measurements $N_{M}$ per unitary and b) the number of unitaries; in a) $N_U=1000$,  $\hilbertn=64$ (dots) and $\hilbertn=256$ (triangles), in b) $N_M=\infty$, $\hilbertn=256$.  c)  Average statistical
		error of an estimation of $p_2=\tr[]{\rho_A^2}$ as a function of the purity of the input state $p_2$. 
		Different colors correspond to different number of measurements $N_M$ per unitary. $\hilbertn=256$, $N_U=100$.
		d) Average statistical error of an estimation of $p_2=\tr[]{\rho_A^2}$ as a function of the number of measurements $N_{M}$ per unitary for observables with different number of outcomes $N_\mathcal{O}=\hilbertn,\hilbertn/2,\hilbertn/4,\hilbertn/8$, with $\hilbertn=256$, $N_U=1000$.
		 In all panels, unitaries were drawn directly from the CUE \cite{Mezzadri2006}. The black solid lines correspond to the analytical expressions given in the text.
		 }
	\label{fig:errors} 
\end{figure}

\textit{Finite number of measurements per unitary} --
Given a fixed random unitary $U_A$, we discuss first the  faithful estimation of the probability  $\mathrm{P} \equiv \tr[]{U_A \rho_A U_A^\dagger \mathcal{P^\mathcal{O}_\vec{s}}}$ of an outcome $\mathbf{s}$ and its powers $P^n$ ($n\in \mathbb{N}$) using $N_M$ measurements.  In an experiment,
$\mathrm{P}$  is obtained  by performing $N_M$  measurements on the state  $U_A \rho_A U_A^\dagger $ and  dividing the  number of measurements, which have yield the outcome $\vec{s}$, by $N_M$. 
Mathematically, this is described by  a random variable (an estimator) $\widetilde {P}$
\begin{align}
\widetilde {P} \equiv \frac{1}{N_M} \sum_{l=1}^{N_M} 1_\mathbf{s}\left({\widetilde X_l}\right)  \stackrel{d}{=} \frac{\widetilde B( \mathrm{P},N_M)}{N_M} \; .
\end{align}
Here, $\widetilde X_l$ is a random variable describing the $l-$th measurement and $1_\mathbf{s}\left( \widetilde X_l\right) = 1 $ if $\widetilde X_l=\mathbf{s}$ and $1_\mathbf{s}\left( \widetilde X_l\right) = 0  $ if $\widetilde X_l\neq\mathbf{s}$. 
$\widetilde{B}(\mathrm{P},N_M)$  denotes a  random variable distributed according to Binomial distribution with probability per trial $\mathrm{P}$ and number of trials $N_M$ \cite{Wadsworth1960}. 
We note that  the estimator $\widetilde {P}$ is faithful, $\mathbb{E}\left[\widetilde {P}\right] = \mathrm{P}$. In contrast, the
 simple power $\widetilde {P}^n$  is a biased estimator of $\mathrm{P}^n$ for $n>1$, it overestimates for finite $N_M$, 
$
\mathbb{E}\left[ \widetilde  {P} ^n \right] - \mathrm{P}^n = \mathcal{O} \left( {\mathrm{P}}/{N_M^{n-1}} \right)
$. 
A unique faithful estimator $\widetilde {P_n} $ is constructed  using an ansatz  $\widetilde {P_n} = \sum_{k=0}^{n} \;  \alpha_k \; \widetilde {P} ^k$ and the defining condition
$\mathbb{E}\left[\widetilde {P_n}\right] \stackrel{!}{=} \mathrm{P}^n $. For example, one finds $\widetilde {P_2}= \widetilde  {P} ( \widetilde  {P} N_M-1 )/ (N_M -1)$ with $\mathbb{E}\left[\widetilde {P_2}\right] = \text{P}^2$. We use these unbiased estimators to simulate numerically an experiment with a finite number of measurements. 

 The average statistical error of an estimation of $\mathrm{P}^n$ using  $\widetilde {P_n} $ is given by its  standard error $\sigma\left[ \widetilde {P_n} \right]$. With the relation to the Binomial distribution, one finds for  $1 \ll N_M \ll 1/\mathrm{P}\approx \hilbertn$
\begin{align}
\sigma\left[ \widetilde {P_n} \right]  = \sqrt{n!} \left( \frac{\mathrm{P}}{N_M} \right)^{\frac{n}{2}}  \left(  1+ \mathcal{O}\left(\frac{1}{N_M} , \mathrm{P}N_M\right) \right) \; .
\label{eq:errs}
\end{align}

\textit{Finite number of unitaries} -- 
The estimation of $\mathrm{P}^n_{U_{A,l}}\equiv\tr{U_{A,l} \rho_A U_{A,l}^\dagger \mathcal{P^\mathcal{O}_\vec{s}}}^n$  ($l=1,\dots, N_U$)  is performed for $N_U$ different random unitaries $U_{A,l}$.  The resulting estimates $(\mathrm{P}_{U_{A,l}}^n)_e$, each  afflicted with an error of the order $\sigma\left[ \widetilde {P_n} \right]$, are subsequently averaged to obtain
$\left\langle \mathrm{P}^n \right \rangle_e=1/N_U \sum_{l=1}^{N_U} (\mathrm{P}^n_{U_{A,l}})_e$. This is an approximation of the average $\left\langle \mathrm{P}^n\right \rangle$  of the exact probabilities   over  the entire CUE. 
Its average statistical error  is  in the limit $N_U\gg1$ determined by the central limit theorem and  given by
\begin{align}
|\left\langle \mathrm{P}^n \right \rangle_e -\left\langle \mathrm{P}^n\right \rangle| \sim  \frac{1}{\sqrt{N_U} } \left(\sigma[\tilde{P}_n] + C_n  \right)
\label{eq:erru}
\end{align}    
It consists of two parts. The former, $ \sigma[\tilde{P}_n]$, has been already discussed and is caused by the finite number of measurements $N_M$ per random unitary.  The latter $C_n$ originates solely  from  the finite number of random unitaries. As noted also in Ref.\ \cite{VanEnk2012} for the case $n=2$,  it is state dependent, $C_n=C_n(\tr[]{\rho_A^n})$,  largest for pure states and given by
\begin{align}
	 C_n  &=  \left( \probs[2n]{\mathbf{s}} -\probs[n]{\mathbf{s}} \right)^{1/2} \nonumber \\&\stackrel{\tr[]{\rho_A^n}=1}{=}  \frac{ n!}{{\hilbertn}^n}   \left( \left[\binom{2n}{n} -1 \right] ^{\frac{1}{2}}  + \mathcal{O}\left(\frac{1}{\hilbertn}\right) \right) \; .
\end{align}
As a function of $\tr[]{\rho_A^n}$, $C_n$ decreases polynomially.  We discuss  its influence  on the total error of an estimation of $\tr[]{\rho_A^n}$ in detail below [see also Fig.~\ref{fig:errors}, panel (c)].

\textit{Total error} --
Finally, an estimate  $(p_n)_e$ is calculated using  a (non-linear) function of all estimated $\left\langle P^k \right \rangle_e$ with $k\leq n$ (see Sec.\ \ref{sec:protocol}). Since, their statistical errors  [equation \eqref{eq:erru}] grow with $k$,  we can estimate the scaling of  total statistical error of $\left(p_n \right)_e$ from the error of the highest order polynomial   $\left\langle \text{P}^n \right \rangle_e$.
To account for the fact that the statistical errors of lower order polynomials are not independent, we introduce further adjustable prefactors. We  find  that the numerical data ($n=2,3,4$) presented in Fig.~\ref{fig:errors}, panels (a) and (b) is  well described by
\begin{align}
| \left(p_n \right)_e - p_n| & \sim \frac{ 1}{\sqrt{N_U  \hilbertn }}  \left( C^\prime_n + B_n \sum_{k=0}^{k<n/2}\left( \frac{\hilbertn }{N_M} \right) ^{\frac{n}{2}-k} \right) \; . \label{eq:pn1}
\end{align}
For $n=2,3,4$, we find 
$C^\prime_n \approx C_n/3(n-1)!$ and $B_n \approx \sqrt{n!}$ which increases exponentially with $n$. The additional global factor $1/\sqrt{\hilbertn}$ in the equation  \eqref{eq:pn1}, compared to \eqref{eq:erru},  originates from the fact  that $(p_n)_e$ is calculated for each possible outcome $\vec{s}$ separately and an average over all $N_\mathcal{O}=\hilbertn$ outcomes has been taken. 

To estimate  $p_n$ up to a fixed error $\epsilon$, one needs hence to perform $N_M\sim \mathcal{N}_A ^{\frac{n-1}{2}}$ measurements per unitary and average over $N_U \sim B^2_n/\epsilon ^2$ unitaries.
While the amount of necessary experimental resources increases thus exponentially with the order of the R{\'e}nyi entropy $n$ and system size, the measurement of low order R{\'e}nyi entropies $n=2,3,4$ in systems of moderate size is accessible, as also exemplified  by the various physical examples presented in Ref.~\cite{Elben2018}.
As a concrete example, for a measurment of the purity of $L=14$ spins, $\hilbertn=2^L=16384$, we find a statistical error of $5\%$ using $N_U=100$ unitaries and $N_M=500$ measurements per unitary.

\textit{Influence of the input state} --
We discuss now the influence of the input state on statistical errors. 
Hereby, we note that the random unitaries in the protocol are drawn from an ensemble which is invariant under unitary transformation (the CUE or  an unitary $n$-design). By definition, average statistical errors of $(p_n)_e$ can hence only depend on properties of  $\rho_A$ which are invariant under unitary transformations, i.e.\ on the functionals $p_n$. Accordingly, we show 
in Fig.\ \ref{fig:errors}, panel (c), the average statistical error of the estimated purity $(p_2)_e$  as a function of the purity of the input state $p_2$. In the case $N_M\rightarrow \infty$ (dark blue squares), the error due to the finite number of unitaries, which is proportional to $C_2$ dominates. It is decreasing polynomially with $p_2$ towards zero   for a completely mixed state. However, in the experimentally most relevant scenario, $N_M \lesssim \hilbertn$, we find independence of the initial state.

\textit{Influence of the chosen observable} --
As estimates $(p_n)_e$ can be inferred from random measurements of any observable (see Sec.~\ref{sec:protocol}), we finally discuss  the influence of the chosen observable $\mathcal{O}$ on the statistical error. We note that $(p_n)_e$
is calculated  for each measurement  outcome $\vec{s}$  separately [see equation \eqref{eq:higher}].  Since in an experiment, the probabilities $\probsU{\vec{s}}$ are measured simultaneously for all possible measurement outcomes $\vec{s}$,  an average can be taken without further experimental effort. This reduces the error by approximately a factor $1/\sqrt{N_\mathcal{O}}$ where $N_\mathcal{O}$ is the number of measurement outcomes. Hence, observables with the maximal number of outcomes $N_\mathcal{O}=\hilbertn$ are favorable. This is confirmed numerically, in Fig.\ \ref{fig:errors}, panel (d), where the error of the purity inferred from random measurements of  observables with various numbers of possible outcomes are shown. Varying $N_\mathcal{O}$ lead for the statistical error of the purity to a simple rescaling $
 | \left(p_2 \right)_e - p_2| \sim \left( C_2^\prime + B_2\frac{\hilbertn}{ N_M} \right) /\sqrt{N_U N_\mathcal{O}} $.

\subsection{Imperfections}
 Our measurement scheme relies on the ability (i) to evolve coherently quantum systems and (ii) to reproduce this evolution with high-fidelity (in order to access observables via projective measurements) (iii) to perform these projective measurements with high fidelity. Thus one must carefully investigate the sensitivity to decoherence and imperfections (as familiar from atomic quantum simulation~\cite{Bloch2012}). 
 Below, we quantify these requirements and find that they are compatible with current AMO setups.

\subsubsection{Decoherence}
In Sec.~\ref{sec:quenches}, we estimate the minimum time $T_\mathrm{tot}$, which is required to generate approximate $n$-designs via random quenches.
The goal of this section is to study the effect of decoherence, acting on the system during $T_\mathrm{tot}$, focusing on the measurement of purities ($n=2$).

We describe the time evolution of the sub-system $A$ density matrix via a Markovian master equation
\begin{eqnarray}
\dot \rho_A = \left(\mathcal{L}_D+\mathcal{L}_U \right) \rho_A, 
\end{eqnarray}
where $\mathcal{L}_D$ is the Liouvillian corresponding to coupling to a Markovian bath and $\mathcal{L}_U$ corresponds to the unitary part of the evolution.

We now assume that the rate $\gamma$ associated with the decoherence processes are small compared to the frequency scales of the unitary evolution (i.e the eigenvalues of $H_A^j$) and use a Trotter approximation to describe the final state
\begin{eqnarray}
\rho_A^f = \mathcal{D} \mathcal{U} \mathcal{D} \rho_A, \label{eq:rhodeco}
\end{eqnarray}
with the unitary operator $\mathcal{U}(\rho)= U_A \rho U_A^\dagger$ and the Liouvillian evolution operator $\mathcal{D}=e^{\mathcal{L}_D T_\mathrm{tot}/2}$.
This approximation allows to obtain the first order corrections to $\rho_A^f$ in $\gamma T_\mathrm{tot}$, with respect to the ideal unitary case $\gamma=0$. From Eq.~\eqref{eq:rhodeco}, we notice that decoherence mechanisms affect in this approximation the state of the system before and after the unitary $U_A$ is applied. Note that the first contribution, acting directly on $\rho_A$, implies that the error cannot be corrected without prior knowledge of the input state $\rho_A$.
 \begin{figure}[t]
	\includegraphics[width=0.95\columnwidth]{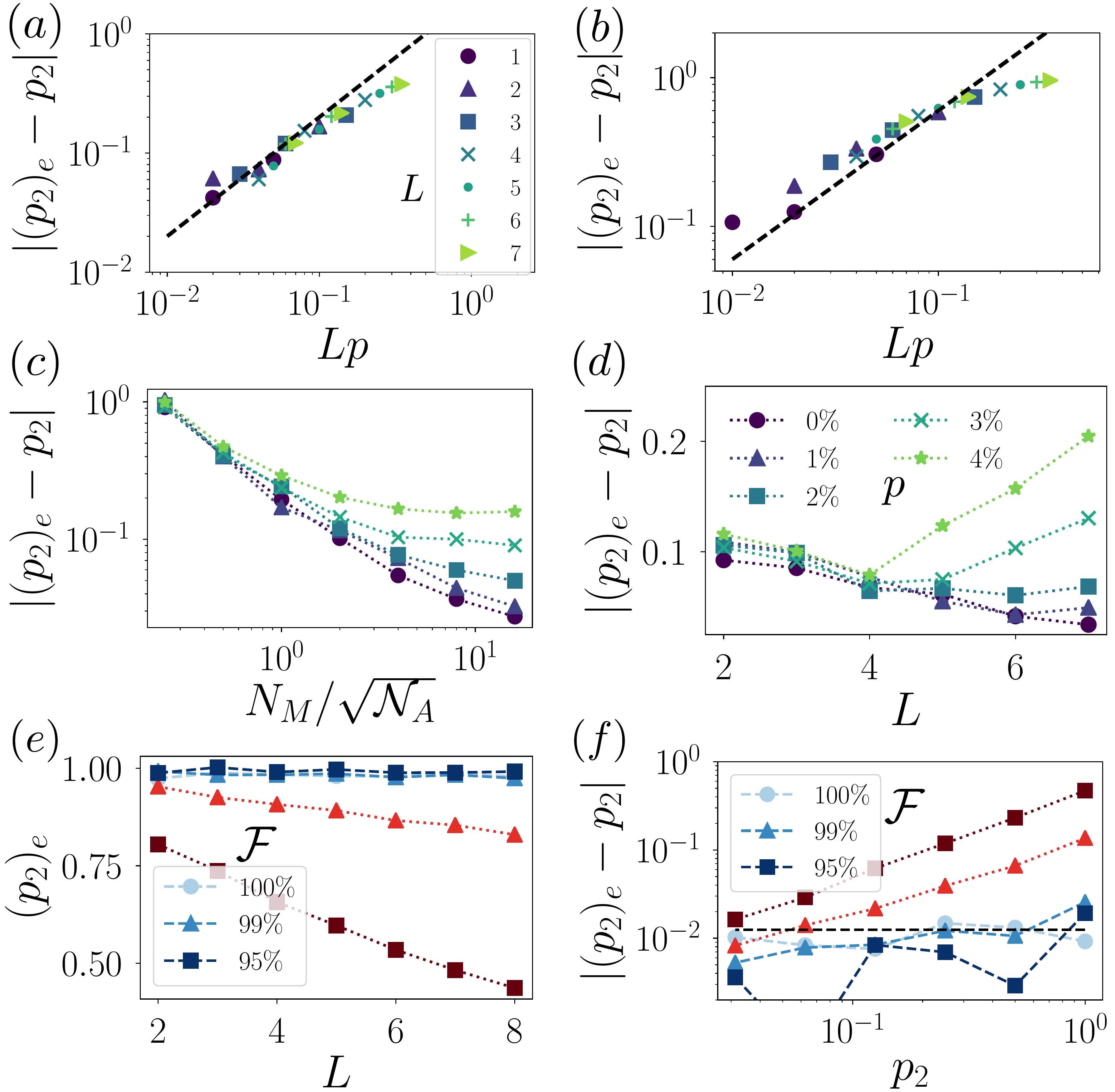}
	\caption{ {\it Imperfections of the measurement scheme}.
		(a-b) Decoherence effects in the measurement of R{\'e}nyi entropies. Relative error of the estimated purity when applying the protocol for a set of $L=1,..,7$ spins with jump probability $p=0.01,0.02,0.05$, after the dissipative evolution given by Eqs.~\eqref{eq:rhodeco},\eqref{eq:channels}. For both dissipative channels, dephasing (a) and depolarizing (b), the error scales linearly with the number of atoms $L$ and the probability $p$.
		In panel (a), respectively (b), the black dashed line represents the total jump probability $2Lp$ (resp $6Lp$). For each point represented,  we consider one set of $N_U=500$ random unitaries, drawn from the CUE.
		(c-d) Error of the estimated purity with imperfectly reproduced disorder patterns (variation of order  $p\delta$  for each projective measurement) created by $\eta=2L$ quenches  of time $T=1/J$ and disorder strength $\delta=J$, (c) as a function of $N_M$ for $L=6$ (d) as a function of $L$ for $N_M=\mathcal{N}_A$. For both panels, we consider $N_U=100$. (e-f) Influence of finite detection fidelity $\mathcal{F}$ on the estimation of the purity, with  uncorrected (red with dotted lines) and corrected (blue with dashed lines) estimators. Shown are (e)  $(p_2)_e$ of a pure antiferromagnetic test state as function of the number of spins $L$ and (f) the error of the estimated purity as a function of the purity $p_2$ of the test state for $L=6$ spins. In both panels, $N_U=100$, $N_M=\hilbertn$.
		\label{fig:deco}}
\end{figure}

For our numerical analysis, we consider the dephasing and fully depolarizing channels with Liouvillan operator~\cite{Preskill2001}
\begin{eqnarray}
\mathcal{D}&=& \prod_{i=1}^L \mathcal{D}_i \label{eq:channels} \\ 
\mathcal{D}_i(\rho)&=& (1-p)\rho+p\sigma_i^z \rho \sigma_i^z \nonumber \quad (\mathrm{dephasing})\\
\mathcal{D}_i(\rho)&=&(1-3p)\rho+p\sum_{\alpha=x,y,z}\sigma_i^\alpha \rho \sigma_i^\alpha \nonumber \quad (\mathrm{depolarizing}),
\end{eqnarray}
with $p= \gamma T_\mathrm{tot}/2$, where $\gamma$ is the decoherence rate. This decomposition in terms of Kraus operators is valid in first order in $p \ll 1$. We use a system of $L$ spins $1/2$ and the pure state $\rho_A=\ket{\psi_\text{ME}}\bra{\psi_\text{ME}}$, with 
\mbox{$\ket{\psi_\text{ME}}=(\ket{\uparrow}^L+\ket{\downarrow}^L)\sqrt{2}$},  which is fragile against both channels. We finally assume that the random quenches generate an exact $2$-design and sample unitaries $U_A$ directly from the CUE.
The relative error of the estimated purity is shown in Fig.~\ref{fig:deco}(a-b). For the dephasing channel [panel (a)], we notice that the error scales linearly with the decoherence probability $p$ and the number of atoms $L$. It approximately corresponds to the total probability $2pL$ (shown as dashed lines) of  having, in the quantum trajectory picture, a quantum jump. The same observation applies to the depolarizing channel, with total probability $6pL$ [panel (b)].

In summary, decoherence mechanisms lead to a linear scaling of errors with respect to the system size $L$ and the local probability of errors $p$. Given that for a system size $L$, the required time to create unitaries is $T_\mathrm{tot} \sim L J^{-1}$, where $J$ here refers to the Hamiltonian timescale ($J\to t_F$ for the FH model). This translates to a condition for the decoherence rate  $\gamma \ll J/L^2$, which can be satisfied in current AMO setups with moderate system sizes.

\subsubsection{Limited reproducibility of generated unitaries}

To estimate for a fixed  random unitary $U_A$ the probability $\text{P}(\mathbf{s})$ of an output $\mathbf{s}$, a set of $N_M\gg1$ projective measurements have to be performed on the same state $\rho_A^f=U_A \rho_A U_A^\dagger$. Hence, for each measurement, the same random unitary $U_A$ has to be created, via the same random quench dynamics.  Here, we discuss the influence of  imperfect reproducibility of the random unitaries (i.e.\ the disorder patterns implementing random quenches) on statistical errors.

To be specific, we consider the Quantum Ising model with $u=z$ and $\alpha=6$ (Rydberg atoms), and $\rho_A=\ket{\uparrow..\uparrow}\bra{\uparrow..\uparrow}$ as test state.
To create a random unitary $U_A$, $\eta$ disorder patterns $\delta_{i}^{j}$ ($i=1,\dots,L$) are drawn from a normal distribution with standard deviation $\delta=\Omega=C_6/a^6$ where $a$ is the lattice spacing. 

We now assume that for each of the $m=1,..,N_M$ projective measurements, these disorder patterns, and thus the random unitary $U_A$, are only imperfectly reproduced.  To simulate this numerically, we add for each measurement additional disorder patterns $\tilde\delta_{i}^{j,m}$ to $H_A^j$ drawn from a normal distribution with standard deviation $p\delta$ ($p\ll 1$).   The  resulting statistical error of the purity [shown in Fig.\ \ref{fig:deco}, panel (c)], does not decrease as a function of the number of measurements $N_M$ towards $1/\sqrt{N_U \hilbertn}$ but converges to a higher value, set by the size of the error $p$. For   $N_M=\mathcal{N}_A$ measurements and $\eta=2L$, we observe that the error increases approximately linearly with $L$ [panel (d)], due to the larger number $\eta \sim L$ of quenches to create a random unitary. 

In summary, assuming a reproducibility of disorder patterns with an error $p$ of a few percent, and moderate system sizes, the estimation of the purity can be obtained with good accuracy.

\subsubsection{Finite detection fidelity}

In this section, we estimate the effect of a finite detection fidelity  $\mathcal{F}$ on the estimation of R\'enyi entropies, and show that it can be taken into account in the protocol. We consider here a spin $1/2$-system with $L$ spins, where for each projective measurement the configuration of each spin can be associated with the wrong state with probability $p=1-\mathcal{F}$. Neglecting other imperfections, the estimated probability $(\text{P}(\mathbf{s}))_e$ is in the limit $N_M \to \infty$ then given by
\begin{align}
	(\text{P}(\mathbf{s}))_e =  \text{P}(\mathbf{s}) (1-p)^L + \sum_{\mathbf{s}'} p^{d(\vec{s},\vec{s}')}  \text{P}(\mathbf{s}'),
\end{align}
where we sum over all spin configurations (basis states) $\vec{s}$ and $d(\vec{s},\vec{s}')$ denotes the Hamming distance between the spin configurations $\vec{s}$ and $\vec{s}'$. Using that  $\langle \text{P}(\mathbf{s}) \text{P}(\mathbf{s}') \rangle \approx (1+\delta_{\vec{s},\vec{s}'} \, p_2)/\hilbertn^2$ for $\mathcal{N_A} \gg 1$ (see section \ref{sec:protocol}), we  find that the measured purity  
\begin{align}
	(p_2)_e = (1-p)^{2L} p_2 + \mathcal{O}(p^2)\label{eq:p2biased}
\end{align}
underestimates its true value $p_2$ to leading order in $p$ by a factor  $(1-p)^{2L}$ which is independent of $p_2$. This implies that for a known detection fidelity $\mathcal{F}=1-p$, we can hence correct for this biased error by the simple substitution  $(p_2)_e \rightarrow (p_2)_e /(1-p)^{2L}$.

These analytical findings are confirmed by the numerical results presented in Fig.~\ref{fig:deco} (e-f). Using the standard estimation, we find that a finite detection fidelity leads to a (biased) error increasing  with system size $L$ (red lines), as predicted by Eq.~\eqref{eq:p2biased}. When using the corrected estimates (blue lines), we obtain an accurate determination of the purity within the statistical error threshold (Sec.~\ref{sec:errors}), independently and without a priori knowledge of $p_2$. We note that this holds in particular for a finite number of measurements $N_M$.

\section{Connection to many-body quantum chaos}\label{sec:chaos}

\newcommand{\IPR}{\text{IPR}}

In Sec.~\ref{sec:quenches}, we use the error of the estimated purity $|(p_2)_e-p_2|$ of various states to test the convergence of random unitaries ensembles. In this section, we compare this quantity to other tools  certifying  random unitaries based on random matrix theory~\cite{Guhr1998} and quantum chaos \cite{Haake2010}. We consider the inverse participation ratio (IPR) and the statistics of the eigenvalues (eigenphases) of the created unitary operators. In particular, we identify the  transition from many-body localized ($\eta=1$) to chaotic dynamics ($\eta\gg1$).

Given a unitary $U_A$ with eigenstates $\ket{\phi_\nu}$  and eigenvalues $e^{i\theta_\nu}$ ($\nu=1,\dots, \hilbertn $), the IPR of  a state $\rho_A$
\begin{align}
\IPR (\rho_A) = \sum_{\nu=1}^{\hilbertn}\tr{\rho_A \ketbra{\phi_\nu}{\phi_\nu}}^2
\end{align}
is a measure for the spreading (delocalization) of $\rho_A$ in the eigenbasis of $U_A$. It is maximal for an eigenstate of $U_A$, $ \IPR(\ketbra{\phi_\nu}{\phi_\nu})=1$ for $\nu=1,\dots, \hilbertn $, and minimal for  a completely delocalized state $\ket{\psi}=1/\sqrt{\hilbertn} \sum_\nu \ket{\phi_\nu}$, $ \IPR(\ketbra{\psi}{\psi})=1/\hilbertn$.
In the following, we consider the ensemble averaged  IPR, $\langle \IPR(\rho_A) \rangle$. For  unitary $2$-designs, in particular thus for the CUE, one finds, using results of Ref.~\cite{Ullah1963},
\begin{align}
	\langle \IPR(\rho_A) \rangle&_{\text{CUE}}  =\nonumber \\  \frac{2}{\hilbertn+1} &\left( \vphantom{\left(\rho_A \right)_{\mu}}\tr{\rho_A^2} \right.+ \sum_{\substack{  {\nu, \mu =1}\\ {\text{with}} \\ {\nu> \mu}  } }^{\hilbertn} \left. \left(\rho_A \right)_{\nu}\left(\rho_A \right)_{\mu} \right) \;.
\end{align}
where $\left(\rho_A \right)_{\nu}$ ($\nu=1,\dots,\hilbertn$) denote the eigenvalues of the density matrix $\rho_A$.

\begin{figure}[t]
	\includegraphics[width=0.98\columnwidth]{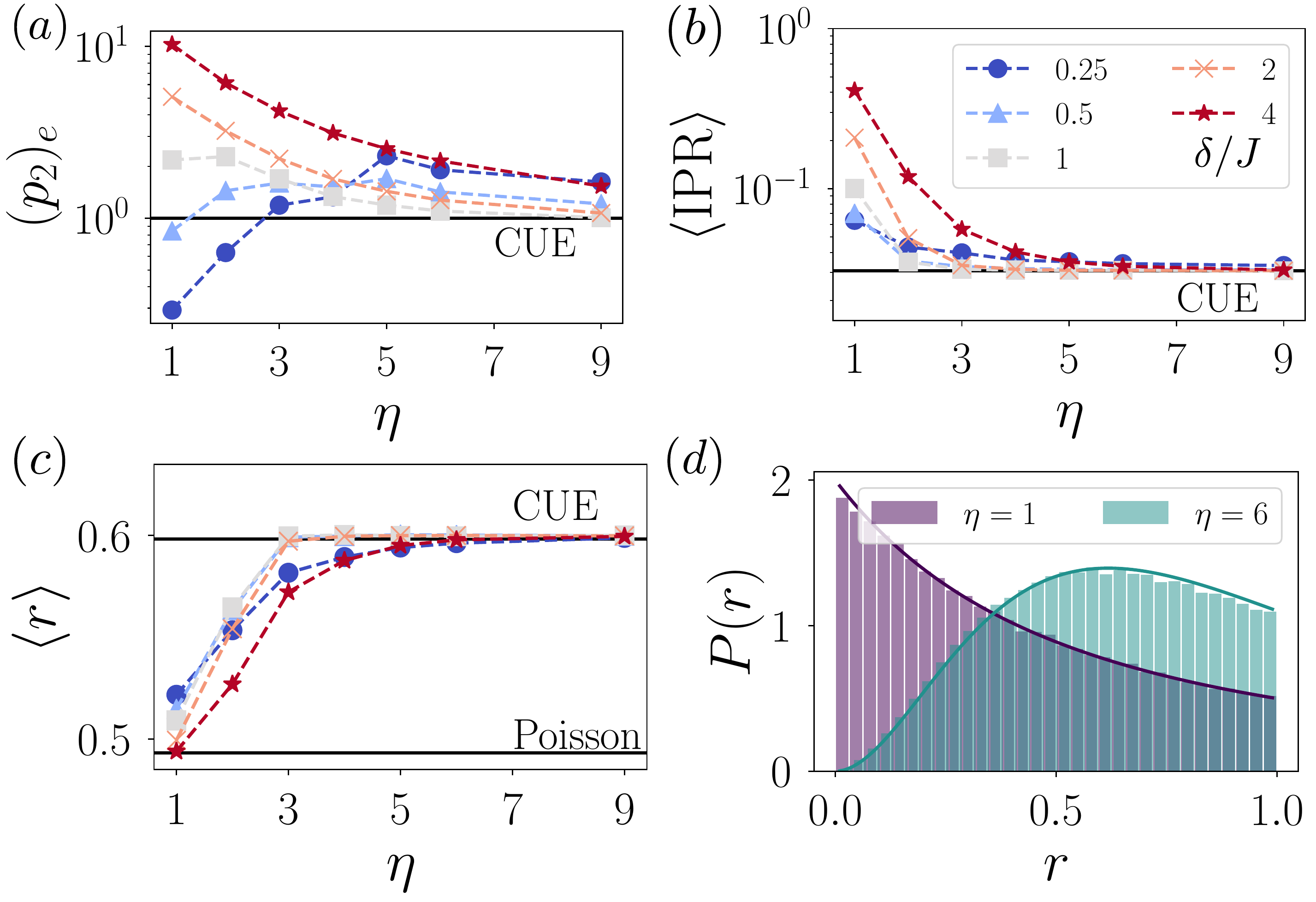}
	\caption{\textit{Comparison of different certification methods}. We use  $N_U=100$  random unitaries  ($N_M=\infty$) created in a Quantum Ising model with $L=6$ spins using random quenches with disorder strength $\delta=0.25,0.5,1,2,4$ [different markers/colors, for the legend, see panel (b)] and the pure test state $\ket{\psi_{AF}} $, to investigate  (a)  the estimation  purity, (b) the ensemble averaged IPR and (c) the mean value $\langle r \rangle$ of the phase spacing statistics $P(r)$, as a function of the depth $\eta$.
		In panel (d), we show $P(r)$ for two values $\eta=1,6$ of depth and disorder strength $\delta/\Omega=4$. Solid lines are theoretical curves for Poisson (purple, resp.\ dark gray) and CUE statistics (green, resp.\ light gray) \cite{DAlessio2014}. 
		\label{fig:Certification}}
\end{figure}

We now consider the phase spacing statistics of the eigenvalues of $U_A$. To avoid the non-trivial unfolding of the spectrum \cite{Haake2010}, we consider the related distribution $P(r)$ of consecutive phase gaps $r$ \cite{DAlessio2014} with
\begin{align}
r = \frac{\min(d_\nu,d_{\nu+1})}{\max(d_\nu,d_{\nu+1})} \in [0,1] \quad \text{with} \;  d_\nu = \theta_{\nu+1} - \theta_\nu, 
\end{align}
which provides information about the characteristic phase repulsion of chaotic quantum systems \cite{Haake2010}.
Here, the phase gaps $d_\nu$ are obtained by first ordering the phases $\theta_\nu \in [-\pi, \pi)$ of the eigenvalues $e^{i\theta_\nu}$ and then calculating the difference between consecutive phases.   Mean values of $r$ for different random matrix ensembles are  $\langle r \rangle_{\text{CUE}}\approx 0.53$ for the CUE and $\langle r \rangle_{\text{POI}}\approx 0.39$ for Poisson distributed phases \cite{DAlessio2014}.
 
In Fig.~\ref{fig:Certification}, we compare the different measures, using random unitaries created with random quenches in a Quantum Ising model with $L=6$ spins, short range interactions $\alpha=6 ($u=z$)$ and  various disorder strengths $\delta$. Quench time and interactions are fixed $JT=1$, $J=\Omega$. We use a pure antiferromagnetic test  state $\ket{\psi_\mathrm{AF}}=\ket{\uparrow \downarrow..\uparrow \downarrow}$. In panels (a-c), we show the estimated purity, the IPR of $\ket{\psi_\mathrm{AF}}$ and the average value of adjacent phase gaps, respectively, as a function of the number of quenches $\eta$. For a wide range of disorder strengths, we find convergence to the respective values expected for the CUE, indicating the convergence of the created ensemble of random unitaries towards the CUE. An optimal speed of convergence is obtained for $\delta \approx J$ for all cases. 

For a single quench $\eta = 1$, we  observe  that the dynamics described  by $U_A=\exp(-iH_A^1 T)$ approach the many-body localized regime with increasing disorder $\delta >J$: In panel (a), the estimation of the purity grows with $\delta$, indicating that the state is less randomized (delocalized) in the entire Hilbert space. This is confirmed in panel (b) by an IPR growing with $\delta$.  Further, we find that $\langle r \rangle$ approaches with growing $\delta$ the Poisson value, which represents a signature of many-body localization (see \cite{DAlessio2014} and references therein).  However, as shown in panel (d),  we find that dynamics become chaotic with increasing  number of quenches, even for  strong disorder $\delta/J=4$. For $\eta=1$ we obtain Poissonian statistics, whereas for $\eta=6$ we observe phase repulsion described by Wigner's surmise~\cite{DAlessio2014}.

In summary, we find that the estimation of the purity of known test states with our measurement protocol provides an experimentally implementable method to test  created ensembles of random unitaries, which is consistent with other theoretical tools like IPR and phase spacing statistics. Beyond the measurement of R\'enyi entropies, we emphasize that random quenches can be used as experimental tool to study the (fast) transition from localized to ergodic dynamics and many-body quantum chaos. 
\section{Conclusion}
In summary, we have extended the analysis presented in Ref.~\cite{Elben2018} for the measurement of R\'enyi entropies in generic lattice models, and discussed different strategies related to the creation of random unitaries via random quenches and the properties of convergence to approximate $n$ designs. Our work opens new possibilities for the experimental study of strongly correlated quantum matter in atomic systems.

\begin{acknowledgments}
We thank the M.~Lukin,  M.~Greiner, M.~Hafezi group members, and J.~Eisert, C.~Roos, P.~Jurcevic, G.~Pagano, W.~Lechner, M.~Baranov, H.~Pichler, P.~Hauke, M.~\L\k{a}cki, and D. ~Hangleiter for  discussions.
The nunerical simulations were performed using the QuTiP library~\cite{Johansson20131234}.
Work in Innsbruck is supported by the ERC Synergy Grant UQUAM and the SFB FoQuS (FWF Project No. F4016-N23). JIC acknowledges support from the ERC grant QUENOCOBA.
\end{acknowledgments}
  
%

\end{document}